\documentclass[12pt]{article}
\usepackage{amssymb,amsmath,latexsym}
\usepackage{graphicx}
\usepackage[final]{pdfpages}
\usepackage{verbatim}

\title{Transport in the random Kronig-Penney model}

\author{Maxim Drabkin$^1$, Werner Kirsch$^2$, Hermann Schulz-Baldes$^1$
\\
\\
{\small $^1$Department Mathematik, Universit\"at Erlangen-N\"urnberg, Germany}
\\
{\small $^2$Fakult\"at f\"ur Mathematik und Informatik, FernUniversit\"at Hagen, Germany}
}

\date{ }

\newtheorem{theo}{Theorem}

\newtheorem{proposi}[theo]{Proposition}
\newtheorem{lemma}[theo]{Lemma}
\newtheorem{coro}[theo]{Corollary}

\newcommand{\CM}{{\mathbb C}}

\newcommand{\NM}{{\mathbb N}}
\newcommand{\RM}{{\mathbb R}}
\newcommand{\SM}{{\mathbb S}}

\newcommand{\ZM}{{\mathbb Z}}

\newcommand{\HM}{{\mathbb H}}
\newcommand{\PM}{{\mathbb P}}

\newcommand{\EE}{{\bf E}}

\newcommand{\Dd}{{\cal D}}

\newcommand{\Oo}{{\cal O}}
\newcommand{\Tr}{\mbox{\rm Tr}}
\newcommand{\Tt}{{\cal T}}

\newcommand{\Nn}{{\cal N}}

\newcommand{\one}{{\bf 1}}

\begin{document}

\maketitle

\begin{abstract}
The Kronig-Penney model with random Dirac potentials on the lattice $\ZM$ has critical energies at which the Lyapunov exponent vanishes and the density of states has a van Hove singularity. This leads to a non-trivial quantum diffusion even though the spectrum is known to be pure-point.


\end{abstract}



\section{Main results}
\label{sec-mainresults}

The Kronig-Penney model describes the motion of a particle in a one-dimensional system with singular potentials. We consider the case of a random $\delta$-potential on the points of the lattice $\ZM$. The Hamiltionian is given by
\begin{equation}
\label{Hamiltonian}
H_{\omega}\;=\; -\,\frac{d^2}{dx^2}\, +\, \sum_{n\in\ZM}v_n \delta_n,
\end{equation}
where $\omega = (v_n)_{n\in\ZM}$ are i.i.d. random variables. We assume that the distribution of the $v_n$'s has  compact support and is nondegenerate, i.~e. is not concentrated in a single point. The precise mathematical meaning of the $\delta$-potential in \eqref{Hamiltonian} is recalled in Section~\ref{sec-setup} below. If the $v_n$'s all have the same sign, the almost sure spectrum of $H_\omega$ consists of an infinite number of bands, while the spectrum contains all positive reals if $0$ is in the support of the distribution of the $v_n$'s \cite{KM} (see Section~\ref{sec-setup} for details). Under rather general assumptions it is known that the spectrum the one-dimensional random Schr\"odinger operators is pure point and that the eigenfunctions are exponentially localized. We are not aware of any publication in which this is actually proved for model \eqref{Hamiltonian}, but the general methods \cite{St} can be applied for this model (details will be given elsewhere, see, however, \cite{HKK} where localization at the bottom of the bottom of the spectrum is proved).

\vspace{.2cm}

The main focus here is on the transport properties of the random Kronig-Penney Hamiltonian. We analyze the growth of the time-averaged $q^{\mbox{\rm\tiny th}}$ moment of the position operator $X$ on $L^2(\RM)$:
\begin{equation}
\label{eq-momentav}
 M_q (T)
\;=\;
\frac{2}{T} \;\int_0^{\infty}dt\;e^{-\frac{2t}{T}}\;e^{\imath H_{\omega}t}|X|^q e^{-\imath H_{\omega}t}
\;,
\qquad
q>0 \;.
\end{equation}
This operator is an integral operator and its integral kernel is denoted by $\langle x|M_q(T)|y\rangle$. The diagonal entries $\langle a|M_q(T)|a\rangle$ can be interpreted as the time-averaged moments of a wave packet initially localized in a Dirac state at $a\in\RM$ (which, of course, is not an element of Hilbert space). The following result shows that for sufficiently large $q$ these moments grow with time $T$, which is certainly not the typical behavior inside the localization regime.

\begin{theo}
\label{theo-transport}
Suppose that $\EE(v)\not = 0$ and let $a\in\RM\setminus\ZM$. For every $\alpha>0$ there is a positive constant $C_\alpha$ such that for $T>1$
\begin{equation}
\label{eq-LB}
\langle a|M_q(T)|a\rangle \;\geq \;C_\alpha \,T^{q(\frac{2}{3}-\frac{5}{3q})-\alpha}
\;,
\qquad
q>\frac{5}{2}\;.
\end{equation}
\end{theo}

The heuristics leading to a non-trivial lower bound \eqref{eq-LB} on the wavepacket spreading is as follows. We will show that the localization length (given by the inverse of the Lyapunov exponent) diverges at certain critical energies $E_l$. At the same time, the density of states diverges at these energies so that there are many such states (see Theorem~\ref{theo-Lyap} below). The quantum motion in these states is more or less ballistic until it reaches the localization length. This decreasing fraction of delocalized states allows to prove the lower bound. Both these heuristics and the implementation of the idea in a proof are quite similar to the random polymer model \cite{JSS}. There is, however, a crucial difference: the transfer matrices at the critical energies $E_l$ are all Jordan blocks in the random Kronig-Penney model, while they are random rotations at the critical energies of a random polymer model. This leads to a completely different behavior of the Lyapunov exponent and ultimately also for the moments in \eqref{eq-LB}. Indeed, in the random polymer model the behavior is $\langle a|M_q(T)|a\rangle\sim T^{q-\frac{1}{2}}$ with lower and upper bounds proved in \cite{JSS} and \cite{JS} respectively. Let us stress though that we do not claim (nor expect) that the lower bound \eqref{eq-LB} is optimal.

\vspace{.2cm}

The proof and above heuristics are based on the vanishing of the Lyapunov exponent $\gamma^E$ at the critical energies $E_l=(\pi l)^2$, $l\in\NM$, which have been known at least since the work of Ishii \cite{Ish}. The Lyapunov exponent is defined in terms of the fundamental solution $\mathcal{T}_\omega^{E}(x,y)\in\mbox{\rm Sl}(2,\RM)$, $x,y\in\RM$, of the first order linear equation in $\RM^2$ associated to \eqref{Hamiltonian}, by the formula
\begin{equation}
\label{eq-Lyapdef}
\gamma^E
\;=\;
\lim_{N\to\infty}\;
\frac{1}{N}\;
\log(\|\mathcal{T}^{E}_\omega(N,0)\,e\|)
\;,
\end{equation}
where $e$ is an arbitrary unit vector in $\RM^2$. The convergence is known to be almost sure \cite{BL}. The next theorem summarizes our main results on the behavior of $\gamma^E$ as well as the integrated density of states $\Nn^E$ in the vicinity of the critical energies. The formal definition of  $\Nn^E$ is recalled in Section~\ref{sec-Lyapunov}.

\begin{theo}
\label{theo-Lyap}
Let $\EE(v)\not =0$. With the positive constants
$$
D_-\;=\; \frac{\EE(v^2-\EE(v)^2)}{16\,\EE(v)E_l}\;,\qquad
D_+\;=\;\left(\frac{\EE(v)}{2E_l}\right)^{\frac{1}{2}}
\;,
$$
one has for $\varepsilon\geq 0$
\begin{equation}
\label{eq-gammamain}
\gamma^{E_l-\varepsilon}
\;=\;
D_-\,\varepsilon\;+\;\Oo(\varepsilon^{\frac{3}{2}})
\;,
\qquad
\gamma^{E_l+\varepsilon}
\;=\;
D_+\,\varepsilon^{\frac{1}{2}}\;+\;\Oo(\varepsilon)
\;,
\end{equation}
and
\begin{equation}
\label{eq-DOSmain}
\Nn^{E_l-\varepsilon}
\;=\;
l-\frac{1}{\pi}\,D_+\,\varepsilon^{\frac{1}{2}}\;+\;\Oo(\varepsilon)
\;.
\end{equation}
\end{theo}

The proof of these formulas is based on perturbation theory for products of random matrices around random Jordan blocks, and uses the techniques of \cite{SS1}. It is well-known that $-\gamma^E+\imath\,\pi\,\Nn^E$ are the boundary value of a Herglotz function $w^z$ which in terms of the Weyl-Titchmarch function $m^z_{\omega,\pm}$ defined below is given by $w^z=\EE(m^z_{\omega,\pm})$. It can be read off Theorem~\ref{theo-Lyap} that
$$
\EE(m^{E_l+z}_{\omega,\pm})
\;=\;
\imath\,\pi\,l\,-\,D_+\,z^{\frac{1}{2}}\,
\,+\,(D_-+\imath B)\,z
\;+\;\Oo(z^{\frac{3}{2}})
\;,
$$
where $B$ is some real constant. Then $\Nn^{E_l+\varepsilon}=
l+\frac{1}{\pi}B\varepsilon+\Oo(\varepsilon^{\frac{3}{2}})$. This constant $B$ vanishes if the support of the distribution of the $v_n$ is strictly positive because in this case $E_l$ is an upper band edge. We do not calculate $B$ below, but believe that it is in principle possible by the techniques presented below (but with considerable effort because the Fokker-Planck operators on $\SM^1$ used in \cite{SS1} become singular).

\vspace{.2cm}

Formula \eqref{eq-DOSmain} shows that the density of states conserves a one-dimensional van Hove singularity at $E_l$ (even though the model is random). This is due to the fact that the random potential cannot move any eigenvalue from below $E_l$ to above $E_l$ and vice versa \cite{KN} (this follows from a basic Sturm-Liouville arguement, see Section~\ref{sec-Lyapunov} for details). The existence of such singularities was already proved in \cite{KN}, but the expansion in \eqref{eq-DOSmain} is new, as are the results on the Lyapunov exponent. For the above random Kronig-Penney model with positive potentials Gredeskul and Pastur \cite{GP} have analyzed the density of states at the lower band edges (this is a Lifshitz tail type regime), see also \cite{KN}.

\vspace{.2cm}

\noindent {\bf Acknowledgements:} W.~K. would like to thank Stas Molchanov and G\"unter Stolz for valuable discussions. The authors are grateful for financial support of the DFG.

\section{Basic analytical set-up}
\label{sec-setup}

\subsection{Definition of the operator}
\label{sec-defop}

Let us begin by considering $H_0=-\partial^2$ as a symmetric operator on the Sobolev space $H^2_0(\RM\setminus\ZM)$, namely $\psi\in L^2(\RM)$ with $\psi',\psi''\in L^2(\RM)$ and such that
$\psi(n)=\psi'(n)=0$ for all $n\in\ZM$. The deficiency spaces $\mbox{\rm Ker}(H_0^*\pm\imath)$ are infinite dimensional and can be explicitly calculated. These deficiency spaces can be seen as direct sum of $2$-dimensional subspaces attached to each point $n\in\ZM$ (that is, the local deficiency indices are $(2,2)$ at each $n$). We are only interested in self-adjoint extensions of $H_0$ that are local in the sense that they only link the left and right limits at one point $n\in\ZM$:
$$
\psi(n_{\pm})\;=\;\lim_{\varepsilon\downarrow0}\;\psi(n\pm\varepsilon),\;
\qquad \psi'(n_{\pm})\;=\;\lim_{\varepsilon\downarrow0}\;\psi'(n\pm\varepsilon)
\;.
$$
The self-adjoint extension corresponding to the $\delta$-potentials in (\ref{Hamiltonian}) is given by
\begin{equation}
\label{eq-bc}
 \begin{pmatrix}
  \psi'(n_+) \\ \psi(n_+)
 \end{pmatrix}\;=\;
\begin{pmatrix}
 1 & v_n \\
 0 & 1
\end{pmatrix}
\begin{pmatrix}
 \psi'(n_-) \\ \psi(n_-)
\end{pmatrix}
\;,
\qquad
n\in\ZM\;.
\end{equation}
This means that the domain $\Dd(H_\omega)$ is given by all functions in $H^2(\RM\setminus\ZM)$ satisfying \eqref{eq-bc}. This gives a precise meaning to the operator in (\ref{Hamiltonian}). If one is interested in $\delta'$-interactions of the form $\sum_n w_n\delta'_n$, then the matrix $\begin{pmatrix}1 & v_n \\ 0 & 1 \end{pmatrix}$ in \eqref{eq-bc} is replaced by $\begin{pmatrix}1 & 0 \\ w_n & 1 \end{pmatrix}$. The main features described in Theorems~\ref{theo-transport} and \ref{theo-Lyap} are also valid in this case. For sake of concreteness, we stick to the model (\ref{Hamiltonian}). Let us point out that it is also possible to consider mixed $\delta$ and $\delta'$ potentials, but then Theorems~\ref{theo-transport} and \ref{theo-Lyap} do not hold (because the matrices describing the boundary conditions are not Jordan blocks in the same basis, which is an essential element in all arguments below).

\vspace{.2cm}

\subsection{Fundamental solutions}
\label{sec-fundsol}

Now we are interested in formal solutions $\psi$ of the Schr\"odinger equation $H_{\omega}\psi = z\psi$ at a complex energy $z$ (possibly not square integrable) which satisfy the boundary conditions \eqref{eq-bc}. As this is a second order differential equation, it is as usual helpful to reformulate it as a system of first order equation for vector valued functions
\begin{equation}
\label{eq-Hamsyscond}
\Psi
\;=\;
\begin{pmatrix}
\psi' \\ \psi
\end{pmatrix}
\;\in\;
H^1(\RM,\CM^2)
\;,
\qquad
\Psi(n+)\;=\;
\begin{pmatrix}
1 & v_n \\ 0 & 1
\end{pmatrix}
\Psi(n-)
\;,
\end{equation}
given by
\begin{equation}
\label{eq-Hamsys}
-\mathcal{J}\,\partial\,\Psi
\;=\;
\begin{pmatrix}
1 & 0 \\ 0 & z
\end{pmatrix}
\Psi
\;,
\qquad
\mathcal{J}\;=\;
\begin{pmatrix}
0 & -1 \\ 1 & 0
\end{pmatrix}
\;.
\end{equation}
Conversely, solving \eqref{eq-Hamsys} with the boundary conditions \eqref{eq-Hamsyscond} gives a solution $\Psi$ the second component of which solves the Schr\"odinger equation $H_{\omega}\psi = z\psi$. These solutions define the transfer matrices, also called fundamental solutions, by the equation
$$
\Psi(x)\;=\;\Tt^z_\omega(x,y)\,\Psi(y)
\;,
$$
as well as the requirement to be right continuous in $x$ and $y$, namely $\Tt^z_\omega(x+,y+)=\Tt^z_\omega(x,y)$. Combined with \eqref{eq-Hamsyscond}, this implies, in particular,
$$
\Tt^z_\omega(n,y)
\;=\;
\begin{pmatrix}
1 & v_n \\ 0 & 1
\end{pmatrix}
\Tt^z_\omega(n-,y)
\;,
\qquad
y<n
\;.
$$
Furthermore, we set for $x<y$,
$$
\mathcal{T}^z_{\omega}(x,y)\;=\; \bigl(\mathcal{T}^z_{\omega}(y,x)\bigr)^{-1}\, .
$$
Then the concatenation identity holds for all $x,y,t\in\RM$:
\begin{equation}
\label{trans}
\mathcal{T}^z_{\omega}(y,x) \;=\; \mathcal{T}^z_{\omega}(y,t)\; \mathcal{T}^z_{\omega}(t,x)\, ,
\end{equation}
Next let us calculate the transfer matrices on an interval in $\RM\setminus\ZM$, for example $[y,x]$, by solving \eqref{eq-Hamsys}:
\begin{equation}
\label{eq-transferraw0}
\Tt^z_\omega(x,y)
\;=\;
\exp\left((x-y)
\begin{pmatrix}
0 & -z \\ 1 & 0
\end{pmatrix}
\right)
\;=\;
\begin{pmatrix}
\cos(z^{\frac{1}{2}}(x-y))  &\ -z^{\frac{1}{2}}\sin(z^{\frac{1}{2}}(x-y)) \\
z^{-\frac{1}{2}}\sin(z^{\frac{1}{2}}(x-y)) & \cos(z^{\frac{1}{2}}(x-y))
\end{pmatrix}
\;.
\end{equation}
Let us note that the second expression is an even function of $z^{\frac{1}{2}}$ so that the choice of square root is irrelevant. For sake of concreteness, we will always choose the principal value though. As $\Tt^z_\omega(x,y)$ is the exponential of a matrix with vanishing trace, it has unit determinant, namely $\Tt^z_\omega(x,y)\in\mbox{\rm Sl}(2,\CM)$. For real energies $E\in\RM$, one, moreover, has $\Tt^E_\omega(x,y)\in\mbox{\rm Sl}(2,\RM)$. It is useful to introduce the notations%
$$
\Tt^z_n\;=\;\Tt^z_\omega(n,n-1)
\;.
$$
According to the above
\begin{equation}
\label{eq-transferraw}
\Tt^z_n \; = \;
\begin{pmatrix}
1 & v_n \\ 0 & 1
\end{pmatrix}
 \begin{pmatrix}
\cos(z^{\frac{1}{2}})  &\ -z^{\frac{1}{2}}\sin(z^{\frac{1}{2}}) \\
z^{-\frac{1}{2}}\sin(z^{\frac{1}{2}}) & \cos(z^{\frac{1}{2}})
\end{pmatrix}
\;.
\end{equation}
Therefore, at $E_l=(\pi l)^2$,
\begin{equation}
\label{eq-transferraw2}
\Tt^{E_l}_n \; = \;
(-1)^l\;\begin{pmatrix}
1 & v_n \\ 0 & 1
\end{pmatrix}
\;.
\end{equation}

\subsection{Basic spectral analysis}
\label{sec-spec}

Here we restrict our attention first to the non-random case of the periodic Hamiltonian $H_v$ defined by setting $v_n = v > 0$. As usual, the spectrum can be read off the transfer matrices $\Tt^E=\Tt^E_n$ at real energy $E=k^2\in\RM$, namely $E\in\sigma(H_v)$ if and only if $|\Tr(\Tt^E)|\leq 2$. The band edges are therefore given by the solutions of the equation
$$
|\Tr(\mathcal{T}^E)| \;= \;\left|\,2\,\cos(k) + \frac{v}{k}\,\sin(k)\,\right| \;=\; 2
\;.
$$
Thus there exist positive constants $c_l(v) \in\RM_+$ such that
$$
\sigma(H_v)\;=\;\bigcup_{l\in\NM}\;\bigl[E_l-c_l(v), E_l\bigr]\;,
\qquad
E_l=(l\pi)^2
$$
In particular, the right band edges are independent of $v$. Thus they are also band edges of the random operator $H_{\omega}$ if the distribution of the $v_n$'s has positive support. Moreover, at these energies the transfer matrices for any periodic approximant are given by
\begin{equation}
\label{edge}
\mathcal{T}^{(l\pi)^2}_{\omega}(n,m)\; = \;(-1)^{l(n-m-1)}
\begin{pmatrix}
 1 & \sum_{j=m+1}^{n}v_j \\
0 & 1
\end{pmatrix}
\;,
\end{equation}
and therefore have a trace of modulus $2$. A little bit more can be said (see \cite{KM} and \cite{GHK}).

\begin{proposi}
\label{prop-spec}
Consider the Kronig-Penney model with positive independent random potentials distributed identically according the probability distribution ${\bf p}$ with compact support in $\RM_+$. Let $v=\inf\;\mbox{\rm supp}({\bf p})$. Then the almost-sure spectrum satisfies
$$
\sigma(H_\omega)\;=\;\sigma(H_v)
\;.
$$
\end{proposi}

\noindent {\bf Proof.}
This follows from a standard Weyl sequence argument using operators with almost constant potential almost equal to $v$.
\hfill $\Box$

\vspace{.2cm}

The following variant is proved similarly.

\begin{proposi}
\label{prop-spec2}
Consider the Kronig-Penney model with independent random potentials distributed identically according the probability distribution ${\bf p}$ with compact support containing $0$. Then the almost-sure spectrum satisfies
$$
[0,\infty)\;\subset\;\sigma(H_\omega)
\;.
$$
\end{proposi}

\subsection{Estimates on the transfer matrices}

Here the focus is on deriving an estimate on the transfer matrices at complex energies $z=E + \kappa$, $\kappa\in\CM$, in terms of estimates at real energies $E>0$. First of all, by analyticity one has
$$
\Tt^z_n\;=\;\Tt^E_n\,+\,\kappa\,R_n
\;,
$$
for some matrix $R_n$ which is uniformly bounded in $n$. Using the concatenation relation (\ref{trans}) one deduces
$$
\mathcal{T}_{\omega}^z(n,\,m)\;=\;\mathcal{T}_{\omega}^{E}(n,m)\,+\,\kappa\;\sum_{j=m+1}^{n} \mathcal{T}_{\omega}^{z}(n,j)\, R_j\, \mathcal{T}_{\omega}^{E}(j-1,m)
\;.
$$
Taking the norm in the above equation, estimating the r.h.s. and taking the supremum over $0\leq m\leq n\leq N$ leads to the following lemma \cite{Sim}.

\begin{lemma}
\label{lem-complexenergies}
Set
$$
c_1\;=\;
\sup_{0\leq m\leq n\leq N} \|\mathcal{T}_{\omega}^E (n,\,m)\| \;,
\qquad
c_2 \;=\; \sup_n \|R_n\|\,.
$$
Then, as long as $|\kappa|\,c_1 \,c_2 \,N < 1$,
$$
\sup_{0\leq m\leq n\leq N} \|\mathcal{T}_{\omega}^{E+\kappa}(n,\,m)\|
\;\leq \;
\frac{c_1}{1-|\kappa|\,c_1 c_2 N }
\;.
$$
\end{lemma}

\subsection{Green functions and Weyl theory}
\label{sec-Green}

At some instances below formulas connecting the transfer matrices to the Green functions will be used. Such a link is established by Weyl theory which is reviewed, {\it e.g.}, in \cite{KS}. For this purpose, one cuts $H_\omega$ into two half-sided operators. This cut is usually done at the orgin $0$, but as the model ~\eqref{Hamiltonian} has a singular potential there, it is more convenient to cut at some other point $a\in(0,1)$. Then it is a basic fact that for every complex energy $z$ with $\Im m(z) >0$ there are two unique $f^z_{\omega,\pm}\in L^2(\RM_\pm)$ with $f^z_{\omega,\pm}(a)=1$ and solving $H_\omega f^z_{\omega,\pm}=z \,f^z_{\omega,\pm}$ on $\RM_\pm$. The $m$-functions are then defined by
$$
m_{\omega,\pm}^z\;=\;
\left(\partial_x\;f_{\omega,\pm}^z\right)(a)\,.
$$
In terms of the transfer matrices, one has
\begin{equation*}
f^z_{\omega,\pm}(x)\;=\;\begin{pmatrix}0 \\ 1\end{pmatrix}^*
\mathcal{T}^z_{\omega}(x,a)^{\pm 1} \begin{pmatrix}\pm\,m_{\omega,\pm}^z \\1\end{pmatrix}\;.
\end{equation*}
It is also known that the resolvent $(z-H_\omega)^{-1}$ is an integral operator with the following jointly continuous kernel
\begin{equation}
\label{eq-Greenformula}
G_\omega^z(x,y)\;=\;
\begin{cases}
f_{\omega,-}^z(x)(m_{\omega,+}^z + m_{\omega,-}^z)^{-1}f_{\omega,+}^z(y)\,,& x\leq y \,,\\
f_{\omega,+}^z(x)(m_{\omega,+}^z + m_{\omega,-}^z)^{-1}f_{\omega,-}^z(y)\,,& y\leq x\,.
\end{cases}
\end{equation}

\section{Lyapunov exponent and DOS at upper band edges}
\label{sec-Lyapunov}

The purpose of this section is to prove Theorem~\ref{theo-Lyap}. Let us begin with the Lyapunov exponent.
It follows from (\ref{edge}) that
\begin{equation*}
 \gamma^{E_l}\; = \;0\;,
\qquad
E_l\;=\;(l\pi)^2
\;.
\end{equation*}
Then \eqref{eq-gammamain} describes how the Lyapunov exponent grows as one enters the spectrum. This will be achieved by a controlled perturbation theory using modified Pr\"ufer variables. This technique also gives access to the density of states.

\subsection{Modified Pr\"ufer variables}

The basic fact motivating the use of modified Pr\"ufer variables is that the Lyapunov exponent defined in \eqref{eq-Lyapdef} can be calculated by
\begin{equation}
\label{eq-Lyaptel}
\gamma^E
\,=\,
\lim_{N\to\infty}\;\EE\,
\frac{1}{N}\;
\log(\|M\mathcal{T}^{E}_\omega(N,0)M^{-1} e\|)
\,=\,
\lim_{N\to\infty}\;\EE\,
\frac{1}{N}\;
\log(\|(M\mathcal{T}^{E}_NM^{-1})\cdots(M\mathcal{T}^{E}_1M^{-1}) e\|)
\,,
\end{equation}
where $M$ is an arbitrary invertible $2\times 2$ matrix which may, moreover, depend on $E$. This matrix $M$ can be chosen later in such a manner that the building blocks $M\mathcal{T}^{E}_nM^{-1}$ are close to some adequately chosen normal form. The choice of $M$ will be made in Section~\ref{sec-choice}.

\vspace{.2cm}

In order to telescop the matrix product in the Lyapunov exponent further below, let us next introduce a random dynamical system on the unit circle. The unit circle is identified with unit vectors in $\RM^2$ via
$$
e_{\theta}\,=\,\begin{pmatrix} \cos{\theta} \\ \sin{\theta} \end{pmatrix} \;,
\qquad
\theta\in [0, \,2\pi)\;.
$$
Then there is a natural action of invertible real $2\times 2$ matrices on the unit circle given by
\begin{equation}
\label{eq-dynamicsS1}
e_{\mathcal{S}_T(\theta)}\,=\,\frac{T e_{\theta}}{\|Te_{\theta}\|}
\;.
\end{equation}
In particular, the map $\mathcal{S}_T$ is invertible and $\mathcal{S}_T^{-1} = \mathcal{S}_{T^{-1}}$. With the notation $u=\binom{1}{-\imath}$, one has
\begin{equation}
\label{expo}
e^{2\imath \mathcal{S}_{T}(\theta)}
\;=\;
\frac{\langle u| T |e_{\theta} \rangle}{\langle\bar{u} | T | e_{\theta} \rangle}
\;.
\end{equation}
In connection with the calculation of the  Lyapunov exponent, one now has to consider the random dynamical system  (Markov process) on the unit circle generated by the random sequence $M\Tt^{E-\varepsilon}_nM^{-1}$ of transfer matrices:
\begin{equation*}
\theta_{n}\;=\;\mathcal{S}_{\varepsilon, n}(\theta_{n-1})
\;,
\end{equation*}
where $\theta_0$ is some initial condition and
\begin{equation}
\label{eq-modPruef}
\mathcal{S}_{\varepsilon, n}\;=\; \mathcal{S}_{M \mathcal{T}^{E_l -\varepsilon}_n M^{-1} }
\;.
\end{equation}
The $\theta_n$ are called the $M$-modified Pr\"ufer phases. If one sets $e_n=e_{\theta_n}$, they are explicitly given by
\begin{equation}
\label{eq-RDS}
e_n\;=\;\frac{M\mathcal{T}^E_n M^{-1} e_{n-1}}{\|M \mathcal{T}^E_n M^{-1} e_{n-1}\|},\qquad \|e_0\|\;=\;1\;.
\end{equation}

\vspace{.2cm}

Coming back to the Lyapunov exponent as given in \eqref{eq-Lyaptel}, one now has
\begin{equation}
\label{eq-LyapBirk}
\gamma^E
\;=\;\lim_{N\rightarrow\infty}\;\frac{1}{N}\;\sum_{n=1}^N\;
\EE\;\log{\Big(\|M^{-1}\mathcal{T}^E_n M e_{n-1}\|\Big)}\:.
\end{equation}
In particular, we now have the Lyapunov exponent given by a Birkhoff sum associated to the random dynamical system \eqref{eq-RDS} (which again converges almost surely so that the expectation in \ref{eq-LyapBirk} may be dropped).

\subsection{The integrated density of states}

The integrated density of states is defined by
\begin{equation*}
 \Nn^E\;=\;\lim_{N\rightarrow\infty}\;\frac{1}{N}\;
\#\{{\rm eigenvalues\;of\; }H_{\omega,N} \leq E\}\;,
\end{equation*}
where $H_{\omega,N}$ is the restriction $H_{\omega}$ to $[0,N]$ with Dirichlet boundary conditions (the Dirac potentials at the boundaries $0$ and $N$ vanish). It is known that the limit defining $\Nn^E$ exists almost surely and is almost surely indpendent of $\omega$. At the critical energies $E_l$, one has
$$
\Nn^{E_l}
\;=\;
l\;.
$$
As pointed out in \cite{KN} this follows from Sturm-Liouville oscillation theory because at energy $E_l$ and volume $N$ one always (for all values of the potential) has an eigenfunction with exactly $Nl$ zeros so that there are exactly $Nl$ eigenvalues below $E_l$. Roughly stated this means that there is no spreading of density of states through each $E_l$ when the distribution of the impurities is changed. Furthermore, Sturm-Liouville oscillation theory allows to calculate the integrated density of states as a rotation number (this can be done at every energy). In the vicinity of the critical  energies, it is, moreover, possible to use the rotation number $R^\varepsilon$ associated to the modified Pr\"ufer phases defined by
\begin{equation}
\label{eq-rotnumber}
R^\varepsilon\;=\;\frac{1}{\pi}\,\lim_{N\rightarrow\infty}\;\frac{1}{N}
\;\EE\;\sum_{n=0}^{N-1}
\;\bigl(\mathcal{S}_{\varepsilon, n}(\theta_{n-1}) - \theta_{n-1}\bigr)
\;.
\end{equation}
In fact, one then has
\begin{equation}
\label{eq-rotnumberIDOS}
\Nn^{E_l-\varepsilon}
\;=\;
l\;+\;
R^\varepsilon
\;.
\end{equation}
%

\subsection{Choice of the basis change in the elliptic regime}
\label{sec-choice}

Close to the critical energy $E_l=(\pi l)^2$, the transfer matrix given by \eqref{eq-transferraw} is close to a Jordan block. Let us begin by expanding $\mathcal{T}^{E_l -\varepsilon}_n$ in $\varepsilon>0$. For sake of simplicity, let us throughout assume that $l$ is even. As $(E_l-\varepsilon)^{\frac{1}{2}}=\pi l - \frac{\varepsilon}{2\pi l} + \Oo(\varepsilon^2)$, one finds

\begin{equation}
\label{eq-transferexpa}
\mathcal{T}^{E_l -\varepsilon}_n\;=\;
\begin{pmatrix}
1 & v_n \\
0 & 1
\end{pmatrix}
\left[\,
\begin{pmatrix}
1 & 0 \\
0 & 1
\end{pmatrix}
\,+\,
\varepsilon
\begin{pmatrix}
0 & \frac{1}{2} \\
-\,\frac{1}{2E_l} & 0
\end{pmatrix}
\,+\,\mathcal{O}(\varepsilon^2)\,
\right]
\;.
\end{equation}
To lowest order, namely for $\varepsilon=0$, the dynamics \eqref{eq-dynamicsS1} of the Jordan blocks have a unique, common, globally attractive fixed point $\theta=0$ which, however, is not stable. Therefore the orbit is close to $\theta=0$ most of the time. This feature is conserved after a small perturbation so that the invariant distribution of the $\theta_n$'s is to lowest order a Dirac peak. Following \cite{DG,SS1}, it is a good idea to blow up  the vicinity of the fixed point in order to detect the effect of disorder and calculate deviations of the invariant measure from the Dirac peak. The blow up will done by conjugating $\mathcal{T}^{E_l -\varepsilon}_n$ with
$$
M_1
\;=\;
\begin{pmatrix}
\varepsilon^{\frac{1}{2}} & 0 \\
0 & 1
\end{pmatrix}
\;.
$$
Indeed,
\begin{equation}
M_1\, \mathcal{T}^{E_l -\varepsilon}_n \,M_1^{-1}
\;=\;
\begin{pmatrix}1 & 0 \\ 0 & 1
\end{pmatrix} \;+\;
\varepsilon^{\frac{1}{2}}\,
\begin{pmatrix}
 0 & v_n  \\
-\frac{1}{2E_l} & 0
\end{pmatrix}
\;-\;
\varepsilon \,
\begin{pmatrix}
\frac{v_n}{2E_l} & 0 \\ 0 & 0
\end{pmatrix}
\;+\;\mathcal{O}(\varepsilon^{\frac{3}{2}})\;,
\end{equation}
which is close to the unit matrix. Next let us center the random variable $v_n$. Set $\bar{v}=\EE(v)$ and $\widetilde{v}_n=v_n - \EE(v)$. Then one can split the $M_1$-modified transfer matrix into a deterministic and a random part:
\begin{equation}
M_1\, \mathcal{T}^{E_l -\varepsilon}_n \,M_1^{-1}
\;=\;
\begin{pmatrix}1-\varepsilon \frac{\bar{v}}{2E_l} & \varepsilon^{\frac{1}{2}}\bar{v} \\ -\varepsilon^{\frac{1}{2}}\frac{1}{2E_l} & 1
\end{pmatrix}
\;+\;
\widetilde{v}_n
\begin{pmatrix}
-\varepsilon\, \frac{1}{2E_l} & \varepsilon^{\frac{1}{2}} \\
0 & 0
\end{pmatrix}
\;+\;\mathcal{O}(\varepsilon^{\frac{3}{2}})\, .
\label{eq-firstconj}
\end{equation}
As long as $\varepsilon \bar{v} \leq4\pi l$, the deterministic summand is up to corrections of order $\varepsilon^{\frac{3}{2}}$ conjugate to a rotation matrix $R_{-\eta\varepsilon^{\frac{1}{2}}}$ by an angle $-\eta\varepsilon^{\frac{1}{2}}$ where
\begin{equation*}
\eta\;=\;\sqrt{\frac{\bar{v}}{2E_l}}
\,.
\end{equation*}
This is what we call an elliptic regime. An equivalent condition for the elliptic regime is that the absolute value of the trace is less than $2$. In this situation it is good to conjugate \eqref{eq-firstconj} again with an adequate matrix $M_2$ in order to bring this rotation into the normal form given by $R_\beta=\begin{pmatrix} \cos(\beta) & -\sin(\beta) \\  \sin(\beta) &  \cos(\beta)\end{pmatrix}$. The good choice turns out to be
\begin{equation*}
M_2\;=\;\begin{pmatrix} \frac{1}{2b} & -b \\  \frac{1}{2b} &  b\end{pmatrix}
\;,
\qquad b=\left(  \frac{\bar{v}   E_l}{2}\right)^{\frac{1}{4}}\;.
\end{equation*}
Indeed, setting $M^\varepsilon=M_2 M_1$ one finds after some algebra
\begin{equation}
M^\varepsilon\mathcal{T}^{E_l -\varepsilon}_n(M^\varepsilon)^{-1}
=
R_{-\eta\,\varepsilon^{\frac{1}{2}}}
\left[\one\,+\,
\varepsilon^{\frac{1}{2}}\, \frac{\widetilde{v}_n}{2\sqrt{2\bar{v} E_l}}
\begin{pmatrix} -1 & 1 \\ -1 & 1 \end{pmatrix}
-
\varepsilon\,\frac{\bar{v}}{4 E_l} \,
\begin{pmatrix} 0 & 1 \\ 1 & 0 \end{pmatrix}
-
\frac{\varepsilon\,\widetilde{v}_n}{2E_l}
\begin{pmatrix} 0 & 1 \\ 1 & 0 \end{pmatrix}
\right]
+\mathcal{O}(\varepsilon^{\frac{3}{2}})
.
\label{eq-transferExpan}
\end{equation}
After this basis change $M^\varepsilon$ the dynamics consists to lowest order $\Oo(\varepsilon^{\frac{1}{2}})$ of a deterministic rotation and a centered random perturbation which thus has a variance of order $\Oo(\varepsilon)$. On the one hand, perturbation theory of the Lyapunov exponent will be based on \eqref{eq-transferExpan}, but on the other it also allows to readily calculate the modified Pr\"ufer dynamics \eqref{eq-modPruef} perturbatively by using \eqref{expo}:
\begin{equation}
\label{eq-Sdyn}
\mathcal{S}_{\varepsilon, n}(\theta)
\;=\;
\theta\,-\,\eta\,\varepsilon^{\frac{1}{2}}\,+\,(\sin(2\theta)-1) \,\frac{\widetilde{v}_n}{2\sqrt{2\bar{v} E_l}}\,\varepsilon^{\frac{1}{2}}
\,+\,\Oo(\varepsilon)\;.
\end{equation}
It is also possible to calculate the terms of $\Oo(\varepsilon)$, but this is not needed here.

\subsection{Perturbative calculation of Birkhoff sums}
\label{sec-invariant}

For the calculation of the Lypapunov exponent and the IDOS by \eqref{eq-LyapBirk} and \eqref{eq-rotnumber} respectively one needs to evaluate Birkhoff sums of the type
$$
I_N(f)\;=\;\EE\;\frac{1}{N}\;\sum_{n=0}^{N-1} \,f(\theta_n)
\;,
\qquad
f\in C(\SM^1)\;.
$$
The following result is taken from \cite{SS1}.

\begin{proposi}
\label{prop-invmeas}
For any $f\in C^1(\SM^1)$, one has
$$
I_N(f)\;=\;\int_0^{2\pi} \frac{d\theta}{2\pi}\; f(\theta)
\; + \;
\mathcal{O}(\varepsilon^{\frac{1}{2}},\,(N\varepsilon^{\frac{1}{2}})^{-1})
\;.
$$
\end{proposi}

\noindent {\bf Proof.} Because $I_N(f)=c+I_N(f-c)$ for $c=\int_0^{2\pi} \frac{d\theta}{2\pi}\, f(\theta)$, one may assume that $\int_0^{2\pi} d\theta f(\theta)=0$. Then $f$ has an antiderivative $F\in C^2(\SM^1)$. Using a Taylor expansion,
$$
F(\theta_n)
\;=\;
F(\mathcal{S}_{\varepsilon,\,n}(\theta_{n-1}))
\;=\;
F(\theta_{n-1})+f(\theta_{n-1})\,\varepsilon^{\frac{1}{2}}\,
\left(
(\sin(2\theta_{n-1})-1)\, \frac{\widetilde{v}_n}{2\sqrt{2\bar{v} E_l}}-\eta
\right)
\;+\; \Oo(\varepsilon)
\;.
$$
As $\widetilde{v}_n$ is centered and independent of $\theta_{n-1}$, taking the expectation and  summing over $n$ shows:
$$
I_N(F)\;=\;I_N(F) \,-\, \varepsilon^{\frac{1}{2}}\,\eta\, I_N(f)\; +\; \Oo(\varepsilon,\, N^{-1})\;.
$$
Dividing by $\varepsilon^{\frac{1}{2}}\,\eta$ finishes the proof.
\hfill $\Box$

\subsection{Calculation of the Lyapunov exponent}
\label{sec-proof}

Here we conclude the controlled perturbation theory for the Lyapunov exponent using the formula \eqref{eq-LyapBirk}. In view of \eqref{eq-LyapBirk}, it is convenient to introduce the auxiliary random variables
\begin{equation*}
 \gamma_n\;=\;\log\bigl(\|M^\varepsilon\mathcal{T}^{E_l-\varepsilon}_n(M^\varepsilon)^{-1}e_{\theta_{n-1}}\|\bigr)\;.
\end{equation*}
Using \eqref{eq-transferExpan}, one finds
\begin{eqnarray}
\gamma_n
& = &
-\;\frac{\widetilde{v}_n}{2\sqrt{2\bar{v}E_l}}\;\cos(2\theta_{n-1})\;\varepsilon^{\frac{1}{2}}
\;+\;
\frac{\widetilde{v}_n}{2E_l}\,\sin(2\theta_{n-1})\,\varepsilon
\nonumber
\\
&  & \;+\;
\frac{\bar{v}}{4E_l}
\,\sin(2\theta_{n-1})\,\varepsilon
\;+\;
\frac{\widetilde{v}_n^2}{16\,\bar{v}\, E_l}
\,\bigl(1-2\sin(2\theta_{n-1})-\cos(4\theta_{n-1})\bigr)\,\varepsilon
\;+\;
\Oo(\varepsilon^{\frac{3}{2}})\;.
\label{eq-gamma_n}
\end{eqnarray}
Now $\widetilde{v}_n$ is centered and independent of $\theta_{n-1}$ so that the expectation value of the terms in the first line vanishes. Furthermore, by Proposition~\ref{prop-invmeas}
$$
\lim_{N\to\infty}\;
\EE\;\frac{1}{N}\;\sum_{n=1}^N
\,\sin(2\theta_{n-1})
\;=\;
\Oo(\varepsilon^{\frac{1}{2}})
\;,
$$
and similarly for the Birkhoff sum of $\cos(4\theta_{n-1})$. Therefore, replacing \eqref{eq-gamma_n} into \eqref{eq-LyapBirk} shows
$$
\gamma^{E_l-\varepsilon}
\;=\;
\lim_{N\to\infty}\;
\EE\;\frac{1}{N}\;\sum_{n=1}^N\,\gamma_n
\;=\;
\frac{\EE(\widetilde{v}_n^2)}{16\,\bar{v}\, E_l}
\,\varepsilon
\;+\;
\Oo(\varepsilon^{\frac{3}{2}})\;,
$$
which is already the formula for $\gamma^{E_l-\varepsilon}$ in Theorem~\ref{theo-Lyap}.

\subsection{Calculation of the IDOS}
\label{sec-proof2}

Next let us calculate the rotation number $R^{E_l-\varepsilon}$ defined in \eqref{eq-rotnumber} by using \eqref{eq-Sdyn}. Again using that $\widetilde{v}_n$ is centered, one finds $R^{E_l-\varepsilon}=-\frac{\eta}{\pi}\varepsilon^{\frac{1}{2}}+\Oo(\varepsilon)$. By \eqref{eq-rotnumberIDOS} this shows the formula for the IDOS in Theorem~\ref{theo-Lyap}.

\subsection{Choice of the basis change in the hyperbolic regime}
\label{sec-choicehyp}

In this section we consider the transfer matrices $\mathcal{T}^{E_l +\varepsilon}_n$ with $\varepsilon>0$. Of course, in \eqref{eq-transferexpa} this only leads to a sign change in front of $\varepsilon$. Then the basis change $M_1$ gives instead of \eqref{eq-firstconj}
\begin{equation}
M_1\, \mathcal{T}^{E_l +\varepsilon}_n \,M_1^{-1}
\;=\;
\begin{pmatrix}1+\varepsilon \frac{\bar{v}}{2E_l} & \varepsilon^{\frac{1}{2}}\bar{v} \\ \varepsilon^{\frac{1}{2}}\frac{1}{2E_l} & 1
\end{pmatrix}
\;+\;
\widetilde{v}_n
\begin{pmatrix}
\varepsilon\, \frac{1}{2E_l} & \varepsilon^{\frac{1}{2}} \\
0 & 0
\end{pmatrix}
\;+\;\mathcal{O}(\varepsilon^{\frac{3}{2}})\, .
\label{eq-firstconjhyp}
\end{equation}
The deterministic part now has trace larger than $2$ and is thus conjugate to a hyperbolic matrix. Again using $M_2$, $M^\varepsilon$ and $\eta$ as above, one now finds
\begin{equation}
M^\varepsilon\mathcal{T}^{E_l +\varepsilon}_n(M^\varepsilon)^{-1}
=
\begin{pmatrix} 1-\eta\varepsilon^{\frac{1}{2}} & 0 \\  & 1+\eta\varepsilon^{\frac{1}{2}} \end{pmatrix}
\,+\,
\varepsilon^{\frac{1}{2}}\, \frac{\widetilde{v}_n}{2\sqrt{2\bar{v} E_l}}
\begin{pmatrix} -1 & 1 \\ -1 & 1 \end{pmatrix}
\,+\,
\varepsilon\,\frac{v_n}{4 E_l} \,
\begin{pmatrix} 1 & 1 \\ 1 & 1 \end{pmatrix}
\,+\,\mathcal{O}(\varepsilon^{\frac{3}{2}})
\;.
\label{eq-transferExpanhyp}
\end{equation}
The phase dynamics in this representation is given by
$$
\mathcal{S}_{\varepsilon, n}(\theta)
\;=\;
\theta\,+\,\varepsilon^{\frac{1}{2}}\,\eta\,\sin(2\theta)\,+\,
\varepsilon^{\frac{1}{2}}
\;\frac{\widetilde{v}_n}{2\sqrt{2\bar{v} E_l}}\;
(\sin(2\theta)-1)
\,+\,\Oo(\varepsilon)\;.
$$
It has $\theta=\frac{\pi}{2}$ as stable fixed point (and $\theta=0$ as unstable one), up to errors of order $\Oo(\varepsilon)$. Replacing this information in \eqref{eq-transferExpanhyp} allows to show (by proceeding as in Section~\ref{sec-proof}) that $\gamma^{E_l+\varepsilon}=\eta \varepsilon^{\frac{1}{2}}+\Oo(\varepsilon)$ which is the missing part of Theorem~\ref{theo-Lyap}.

\section{Transport bounds}

The aim of this section is to prove Theorem~\ref{theo-transport}.

\subsection{Set-up for the lower bound}

By the results of Section~\ref{sec-Green} the resolvent is an integral operator with an integral kernel given by \eqref{eq-Greenformula}. From this follows that also the unitary groups of $H_\omega$ have an integral kernel and we will denote all these integral kernels with Dirac notation. Then
$$
\langle a|M_q(T)|a\rangle
\; = \;
\frac{2}{T}\; \int_0^{\infty} dt\;e^{-\frac{2t}{T}}\;
\int dx \, |x |^q\,
\langle a|e^{\imath H_\omega t}| x \rangle \,\langle x|e^{-\imath H_\omega t}|a\rangle
\;.
$$
The point $a$ is used above in the construction of the resolvent, but can be chosen freely so that one obtains the lower bound for all $a\in\RM\setminus\ZM$.
Writing out the appearing matrix elements in spectral representation and using the identity
$$
\int_0^{\infty} dt\; e^{-\frac{2t}{T}}\;e^{\imath E't}\,e^{-\imath E''t}
\;=\;
\int\frac{dE}{2\pi}\;\frac{1}{E'-E-\imath\, T^{-1}}\;\frac{1}{E''-E+\imath \,T^{-1}}
\;,
$$
it follows that
$$
\langle a|M_q(T)|a\rangle
\; = \;
\int dx\,|x|^q\;\frac{1}{T}\;
\EE\,\int\frac{dE}{ \pi}\;
\Big|\langle x|(H_\omega-E - \imath \,T^{-1})^{-1}|a \rangle \Big|^2
\;.
$$
A lower bound is obtained by restricting the energy integral over an interval $[E_l-\varepsilon_0,E_l]$ close to any of the critical energies:
\begin{eqnarray*}
\langle a|M_q(T)|a\rangle
& \geq &
\int dx\,|x|^q\;
\EE\,\int^{E_l}_{E_l-\varepsilon_0}\frac{dE}{ \pi\,T}\;
\Big|\langle x|(H_\omega-E - \imath\, T^{-1})^{-1}|a \rangle \Big|^2
\\
& = &
\int dx\,|x|^q\;
\EE\,\int^{\varepsilon_0}_0\frac{d\varepsilon}{ \pi\,T}\;
\Big|\langle x|(H_\omega-E_l+\varepsilon - \imath\, T^{-1})^{-1}|a \rangle \Big|^2
\\
& = &
\int dx\,|x|^q\;
\EE\,\int^{\varepsilon_0}_0\frac{d\varepsilon}{ \pi\,T}\;
|G_\omega^z(x, a)|^2
\;,
\end{eqnarray*}
where in the last equality we set  $z=E_l-\varepsilon+\imath\, T^{-1}$. By \eqref{eq-Greenformula} one has for $x\geq a$
$$
|G_\omega^z(x, a)|^2
\; = \;
|G^z_\omega(a,a)|^2 \;\Big|
\begin{pmatrix} 0 \\ 1 \end{pmatrix}^*
\mathcal{T}^z_\omega (x, a)
\begin{pmatrix}
m_{\omega,+}^z \\ 1
\end{pmatrix}\Big|^2
\;,
$$
and similarly for $x\leq a$. Let us also note
\begin{equation}
\label{eq-Greendiag}
|G^z_\omega(a,a)|^2\; = \;
\frac{1}{|m^z_{\omega,+}+m^z_{\omega,-}|^2}
\;.
\end{equation}
Replacing thus shows, still with $z=E_l-\varepsilon+\imath\,T^{-1}$, that $\langle a|M_q(T)|a\rangle$ is bounded below by
$$
\int_0^\infty dx
\,x^q  \,\EE\int^{\varepsilon_0}_0
\frac{d\varepsilon}{\pi\,T}\,
|G_\omega^z(a,a)|^2
\left(
\Big|
\begin{pmatrix}
 0 \\ 1
\end{pmatrix}^*
\mathcal{T}^z_\omega (x, a)
\begin{pmatrix}
m_{\omega,+}^z \\ 1
\end{pmatrix}\Big|^2
+
\Big|
\begin{pmatrix}
 0 \\ 1
\end{pmatrix}^*
\mathcal{T}^z_\omega (-x, a)
\begin{pmatrix}
m_{\omega,-}^z \\ 1
\end{pmatrix}\Big|^2
\right)
\;.
$$
Now it may happen that the appearing matrix elements vanish for some $x$, but as the transfer matrices vary as given by \eqref{eq-transferraw0} between the potentials, for two points $x,x'$ within one interval of $\RM\setminus\ZM$ satisfying $x'\geq x+\frac{1}{2}$, one has
\begin{eqnarray*}
& &
\!\!\!\!\!\!\!\!\!\!\!\!\!\!\!\!\!\!\!\!\!\!\!\!\!
\Big|
\begin{pmatrix}
 0 \\ 1
\end{pmatrix}^*
\mathcal{T}^z_\omega (x, a)
\begin{pmatrix}
m_{\omega,+}^z \\ 1
\end{pmatrix}\Big|^2
\;+\;
\Big|
\begin{pmatrix}
 0 \\ 1
\end{pmatrix}^*
\mathcal{T}^z_\omega (x', a)
\begin{pmatrix}
m_{\omega,+}^z \\ 1
\end{pmatrix}\Big|^2
\\
& = &
 \Big|
\begin{pmatrix}
m_{\omega,+}^z \\ 1
\end{pmatrix}^*
\mathcal{T}_\omega^z (x, a)^*
A\,
\mathcal{T}_\omega^z (x, a)
\begin{pmatrix}
m_{\omega,+}^z \\ 1
\end{pmatrix}\Big|^2
\\
& \geq &
C_1
\;
\Big\|
\mathcal{T}_\omega^z (x, a)
\begin{pmatrix}
m_{\omega,+}^z \\ 1
\end{pmatrix}\Big\|^2
\\
& \geq &
C_1\;\frac{|m_{\omega,+}^z|^2+1}{\|\mathcal{T}_\omega^z (x, a)\|^2}
\;,
\end{eqnarray*}
where $C_1$ is a uniform lower bound (for $E=E_l-\varepsilon\in[E_l-\varepsilon_0,E_l]$ and $T\geq 1$) on the positive matrix
$$
A
\; =\;
\begin{pmatrix}
  0 \\ 1
 \end{pmatrix}
 \begin{pmatrix}
  0 \\ 1
 \end{pmatrix}^* \,+\,
\mathcal{T}^z_\omega (x', x)^*
\begin{pmatrix}
  0 \\ 1
 \end{pmatrix}
 \begin{pmatrix}
  0 \\ 1
 \end{pmatrix}^*
\mathcal{T}^z_\omega (x', x)
\;,
$$
and where in the last inequality it was also used that the transfer matrix is in ${\rm Sl}(2,\CM)$ so that its norm coincides with the norm of its inverse. A similar bound holds for $x<0$. Of course, one can also replace $x$ and $x'$ by each other at the cost of another factor. Therefore the above implies
\begin{equation}
\langle a|M_q(T)|a\rangle
\; \geq \;
\frac{C_2}{T}
\;\int_0^N dx
\;x^q \;\EE \int^{\varepsilon_0}_0\, d\varepsilon\,
|G_\omega^z(a,a)|^2
\left(
\,\frac{|m_{\omega,+}^z|^2+1}{\|\mathcal{T}_\omega^z (x, a)\|^2}
\,+\,
\frac{|m_{\omega,-}^z|^2+1}{\|\mathcal{T}_\omega^z (-x, a)\|^2}
\,\right)
\;,
\label{eq-lowerintermed}
\end{equation}
where the upper boundary $N$ on the space variable can be chosen at convenience later on. Actually, both $N$ and  $\varepsilon_0$ will be adequate functions of time $T$. The bound \eqref{eq-lowerintermed} will be the starting point of the conclusion of the argument below.

\vspace{.2cm}

\noindent {\bf Remark} Let us show how one may go over to the $M^\varepsilon$-modified transfer matrices at this point. First of all, $M^\varepsilon=M_2M_1$ satisfies $\|M^\varepsilon\|\leq C_3$ and $\|(M^\varepsilon)^{-1}\|\leq C_3 \varepsilon^{-\frac{1}{2}}$ because $\|M_1\|=1$,  $\|M_1^{-1}\|=\varepsilon^{-\frac{1}{2}}$ and $C_3=\|M_2\|=\|M_2^{-1}\|$. Thus $\langle a|M_q(T)|a\rangle
$ is bounded below by
\begin{equation}
\frac{C_4}{T}
\;\int_0^N dx
\;x^q \;\EE \int^{\varepsilon_0}_0\, d\varepsilon\,\varepsilon\,
|G_\omega^z(a,a)|^2
\left(
\,\frac{|m_{\omega,+}^z|^2+1}{\|M^{\varepsilon}\mathcal{T}_\omega^z (x, a)(M^{\varepsilon})^{-1}\|^2}
\,+\,
\frac{|m_{\omega,-}^z|^2+1}{\|M^{\varepsilon}\mathcal{T}_\omega^z (-x, a)(M^{\varepsilon})^{-1}\|^2}
\,\right)
\;.
\label{eq-lowerintermed2}
\end{equation}
Replacing \eqref{eq-Greendiag} for $|G_\omega^z(a,a)|^2$ one can now attempt to use the deterministic lower bound
\begin{equation}
\label{eq-bounddeterministic}
\frac{|m_{\omega,+}^z|^2+1}{|m_{\omega,+}^z+m_{\omega,-}^z|^2}
\;+\;
\frac{|m_{\omega,-}^z|^2+1}{|m_{\omega,+}^z+m_{\omega,-}^z|^2}
\;\geq\;
\frac{1}{2}\;
\frac{|m_{\omega,+}^z|^2+|m_{\omega,-}^z|^2+2}{|m_{\omega,+}^z|^2+|m_{\omega,-}^z|^2}
\;\geq\;\frac{1}{2}
\;,
\end{equation}
so that either the first or the second summand in \eqref{eq-lowerintermed2} is bounded below by $\frac{1}{4}$. One is then tempted to combine this with Jensen's inequality
$$
\EE\left(\frac{1}{\|M^\varepsilon\mathcal{T}^z_{\omega}(x,a)(M^\varepsilon)^{-1}\|^2}
\right)
\; \geq \;
\exp\left(-\EE\bigl(\log(\|M^\varepsilon\mathcal{T}^z_{\omega}(x,a)(M^\varepsilon)^{-1}\|^2)\bigr)\right)
\;.
$$
Now one just calculates the expectation in the exponent exactly as the Lyapunov exponent was calculated. However, there is a flaw in this argument: one cannot decorrelate the two bounds because there is no control for which $\omega$ the lower bound by $\frac{1}{4}$ holds. Therefore, one needs to show that $\|\mathcal{T}^z_{\omega}(x,a)\|\lesssim e^{x\gamma^z }$ grows as most as indicated by the Lyapunov exponent $\gamma^z$ at least with high probability. This is shown in the next section. An alternative would be to show that the two terms on the l.h.s. of  \eqref{eq-bounddeterministic} are of equal size with high probability. Another way out is to consider the model on the half-line with adequate boundary conditions (not Dirichlet). Then there is only one term and arguing with Jensen as above works.
\hfill $\diamond$

\subsection{Large deviations}

Here it will be shown that the $M^\varepsilon$-modified transfer matrices appearing in \eqref{eq-lowerintermed2} remain small on adequate scales with high probability. For this purpose, let us work first at real energies, neglect the difference from $x$ and $a$ to  the integers, and also insert an arbitrary initial unit vector $e_0\in\RM^2$.  Working with modified Pr\"ufer variables as in Section~\ref{sec-Lyapunov} in order to telescope the matrix product and then using \eqref{eq-gamma_n} shows
\begin{equation}
\log(\|M^\varepsilon \mathcal{T}_\omega^{E_l - \varepsilon}(n,m)(M^\varepsilon )^{-1}e_0\|)
\; = \;
\;\sum_{j=m+1}^{n}\,\gamma_j
\; = \;
\varepsilon^{\frac{1}{2}}\;
\;\sum_{j=m+1}^{n}\;X_j
\;+\;
\Oo\bigl((n-m)\,\varepsilon\bigr)
\;,
\label{eq-randomsum}
\end{equation}
where
$$
X_j\;=\;
-\;\frac{\widetilde{v}_{j}}{2\sqrt{2\bar{v}E_l}}\;\cos(2\,\theta_{j-1})
\;.
$$
These $X_j$  are martingales, namely centered random variables depending on $v_i$, $i\leq j$, but not on the future $v_i$, $i>j$. They also depend on $\theta_0$ and $\varepsilon$. Let us begin with a standard upper bound on the large deviations of sums of the $X_j$.

\begin{lemma}
Let $\alpha>0$.  Set $Z^\varepsilon(n,m)=\sum_{j=m+1}^n X_j$ and
$$
\Omega^\varepsilon_N(\alpha,e_0)
\;=\;
\left\{\omega\in\Omega\,\left|\,\sup_{-N\leq m\leq n\leq N}|Z^\varepsilon(n,m)|\geq N^{\frac{1}{2}+\alpha}\right.\right\}
\;.
$$
Then there is a constant $C_5>0$ such that
$$
\PM(\Omega^\varepsilon_N(\alpha,e_0))
\;\leq\;
C_5\,N^2\,e^{-N^\alpha}
\;.
$$
\end{lemma}

\noindent {\bf Proof}. As the distribution of the potentials $v_j$ is compactly supported, one has  $-c\leq X_j\leq c$ for some $c>0$. Therefore the convexity of the exponential implies
$$
\EE_j(e^{\beta X_j}) \;\leq \; \cosh(\beta c)\; \leq \;e^{\frac{1}{2}(\beta c)^2}
\;,
$$
where $\EE_j$ is the expectation over $v_{j}$ only. For $\lambda>0$ and $\beta>0$, one thus has
\begin{eqnarray*}
 \mathbb{P}\Big(\big\{Z^\varepsilon(n,m)  \geq \lambda\big\}\Big)
 & = &
 \mathbb{P}\Big(\big\{e^{\beta Z^\varepsilon(n,m)} \geq e^{\beta\lambda}\big\}\Big)
 \;\leq \;
e^{-\lambda\beta}\;\EE(e^{\beta Z^\varepsilon(n,m)})
\\
&  = &
e^{-\lambda\beta}\;\EE_{n-1}\;\EE_n(e^{\beta X_n}e^{\beta Z^\varepsilon(n-1,m)})
\; \leq\;
e^{-\lambda\beta}\;e^{\frac{1}{2}(c\beta)^2}\;\EE_{n-1}(e^{\beta Z^\varepsilon({n-1},m)})
\\
& \leq &  e^{-\lambda\beta}\;e^{\frac{1}{2}(n-m)(c\beta)^2}
\;,
\end{eqnarray*}
where the last inequality is obtained iteratively. A similar bound holds for $\lambda<0$ and $\beta<0$. With $\lambda = N^{\frac{1}{2}+\alpha}$ and $\beta = N^{-\frac{1}{2}}$ it follows, as $n\leq N$,
$$
\mathbb{P}\Big(\big\{ |Z^\varepsilon(n,m)|  \geq N^{\frac{1}{2}+\alpha}\big\}\Big)
\; \leq \;
2 \,e^{\frac{1}{2} c^2} \,e^{-N^{\alpha}}
 \;.
$$
Summing over $m$ and $n$ completes the proof.
\hfill $\Box$

\vspace{.2cm}

Combining the lemma with \eqref{eq-randomsum} leads to the following.

\begin{coro}
\label{coro-dev1}
For $\omega\not\in\Omega^\varepsilon_N(\alpha,e_0)$ one has for all $-N\leq m\leq n\leq N$ and some constant $C_{6}$
$$
\log(\|M^\varepsilon \mathcal{T}_\omega^{E_l - \varepsilon}(n,m)(M^\varepsilon )^{-1}e_0\|)
\;\leq\;
N^{\frac{1}{2}+\alpha}\,\varepsilon^{\frac{1}{2}}
\;+\;
C_{6}\,N\,\varepsilon
\;.
$$
\end{coro}

\begin{coro}
\label{coro-dev2}
There is a set $\Omega^\varepsilon_N(\alpha)$ satisfying
\begin{equation}
\label{eq-Ombound}
\PM(\Omega^\varepsilon_N(\alpha))
\;\leq\;
2\,C_5\,N^2\,e^{-N^\alpha}
\;,
\end{equation}
such that for $\omega\not\in\Omega^\varepsilon_N(\alpha)$, $C_{6}$ as above and a constant $C_{7}$ one has for all $-N\leq m\leq n\leq N$
\begin{equation}
\label{eq-dev2}
\|\mathcal{T}_\omega^{E_l - \varepsilon}(n,m)\|
\;\leq\;
C_{7}\;\varepsilon^{-\frac{1}{2}}\;
\exp\left(N^{\frac{1}{2}+\alpha}\,\varepsilon^{\frac{1}{2}}
\;+\;
C_{6}\,N\,\varepsilon
\right)
\;.
\end{equation}
\end{coro}

\noindent {\bf Proof}. In order to control the norms of the transfer matrices it is sufficient to control their action on $2$ initial vectors because  for any $2\times 2$ matrix $A$,
\begin{equation}
\label{eq-norm}
\left\|A\right\|
\;=\;
\sup_{\theta\in [0,\pi)}\left\|Ae_{\theta}\right\|
\;\leq\;
\sqrt{2}\;\max_{\theta =0, \frac{\pi}{2}}\;
\left\|Ae_{\theta}\right\|
\mbox{ . }
\end{equation}
Hence it is sufficient to prove probabilistic bounds on \eqref{eq-randomsum} for two initial conditions, namely one can set $\Omega^\varepsilon_N(\alpha)=\Omega^\varepsilon_N(\alpha,e_0)\cup \Omega^\varepsilon_N(\alpha,e_{\frac{\pi}{2}})$ which hence satisfies \eqref{eq-Ombound} and it follows from Corollary~\ref{coro-dev1} that for $\omega\not\in\Omega^\varepsilon_N(\alpha)$  and $-N\leq m\leq n\leq N$ one has
$$
\log(\|M^\varepsilon \mathcal{T}_\omega^{E_l - \varepsilon}(n,m)(M^\varepsilon )^{-1}\|)
\;\leq\;
\sqrt{2}\,
\left(N^{\frac{1}{2}+\alpha}\,\varepsilon^{\frac{1}{2}}
\;+\;
C_{6}\,N\,\varepsilon
\right)
\;.
$$
Now the result follows from the bounds $\|M^\varepsilon\|\leq C_3$ and $\|(M^\varepsilon)^{-1}\|\leq C_3 \varepsilon^{-\frac{1}{2}}$ derived before equation \ref{eq-lowerintermed2} above.
\hfill $\Box$

\vspace{.2cm}

Finally, we are going to combine Corollary~\ref{coro-dev2} with Lemma~\ref{lem-complexenergies}. The constant $c_2$ in that lemma is equal to the r.h.s. of \eqref{eq-dev2} as long as $\omega\not\in\Omega^\varepsilon_N(\alpha)$. Therefore, as long as
\begin{equation}
\label{eq-kappabound}
|\kappa|
\;\leq \;\frac{1}{2\,c_2\,c_3\,N}
\;=\;
\frac{\varepsilon^{\frac{1}{2}}}{N} \;\frac{1}{2\,c_3\,C_{11}}\;
\exp\left(-N^{\frac{1}{2}+\alpha}\,\varepsilon^{\frac{1}{2}}
\;-\;
C_{10}\,N\,\varepsilon
\right)
\;,
\end{equation}
one has for $\omega\not\in\Omega^\varepsilon_N(\alpha)$ the bound
\begin{equation}
\label{eq-dev3}
\|\mathcal{T}_\omega^{E_l - \varepsilon+\kappa}(n,m)\|
\;\leq\;
2\;C_{11}\;\varepsilon^{-\frac{1}{2}}\;
\exp\left(N^{\frac{1}{2}+\alpha}\,\varepsilon^{\frac{1}{2}}
\;+\;
C_{10}\,N\,\varepsilon
\right)
\;.
\end{equation}
This means that the bound \eqref{eq-dev2} transposes, up to a factor $2$, to a complex neighborhood of energies the size of which is given by \eqref{eq-kappabound}.

\vspace{.2cm}

The aim is now to use the bound \eqref{eq-dev3} in \eqref{eq-lowerintermed} and for that purpose the free parameters $\varepsilon_0$ and $N$ have to be coupled to the time $T$. In order for \eqref{eq-dev3} to be of any use for all $\varepsilon<\varepsilon_0$, we first of all choose
\begin{equation}
\label{eq-epsN}
\varepsilon_0\;=\;N^{-1-2\alpha}
\;.
\end{equation}
Then the exponential factors in both  \eqref{eq-kappabound} and \eqref{eq-dev3} are of order $1$ and \eqref{eq-kappabound} becomes
$$
|\kappa|
\;\leq\;
C_{8}
\,
N^{-\frac{3}{2}-\alpha}
\;,
$$
for some constant $C_{8}$. Now the size of the balls of radius $\varepsilon\Delta=N^{-\frac{3}{2}-\alpha}$ (in complex energy) is much smaller than the interval of size $\varepsilon_0=N^{-1-2\alpha}$ that we want to cover. However, it can be covered by $N^{\frac{1}{2}}$ intervals of size $\varepsilon\Delta$. Thus setting
$$
\Omega_N(\alpha)
\;=\;
\bigcup_{j=1,\ldots,N^{\frac{1}{2}}}
\;
\Omega^{E_l-j\,\varepsilon\Delta}_N(\alpha)
\;,
$$
one still has a set with sufficiently small probability on the complement of which \eqref{eq-dev3} holds uniformly in an adequate set of complex energies. More precisely, we have proved the following.

\begin{proposi}
\label{prop-dev}
Let $\alpha>0$. There exists a set  $\Omega_N(\alpha)$ satisfying
\begin{equation}
\label{eq-Ombound2}
\PM(\Omega_N(\alpha))
\;\leq\;
2\,C_5\,N^{\frac{5}{2}}\,e^{-N^\alpha}
\;,
\end{equation}
such that there are constants $C_{8},C_{9}$ so that for all $\varepsilon\leq N^{-1-2\alpha}$ and $\delta\leq C_{8}N^{-\frac{3}{2}-\alpha}$ one has for $\omega\not\in\Omega_N(\alpha)$
$$
\sup_{-N\leq m\leq n\leq N}
\; \|\mathcal{T}_\omega^{E_l - \varepsilon+\imath\,\delta}(n,m)\|
 \;\leq \;
C_{9}\;\varepsilon^{-\frac{1}{2}}\;
\;.
$$
\end{proposi}

\vspace{.2cm}

We do not expect this result to be optimal, but rather that the same bound holds for a larger set of complex energies given by $\delta\leq C_{8}N^{-1-\alpha}$, but could not obtain such a better estimate.

\subsection{Conclusion of the proof of the lower bound}

In \eqref{eq-lowerintermed}, we now choose $N$ and $\varepsilon_0$ as
$$
N\;=\;(C_{8}\,T)^{\frac{2}{3+2\alpha}}
\;,
\qquad
\varepsilon_0\;=\;N^{-1-2\alpha}\;=\;(C_{8}\,T)^{-\frac{2+4\alpha}{3+2\alpha}}
\;.
$$
Due to Proposition~\ref{prop-dev} one then has
%
$$
\langle a|M_q(T)|a\rangle
\; \geq \;
\frac{C_2}{T}
\;\int_0^N dx
\;x^q \;\EE
\;\chi_{\Omega_N(\alpha)^c}(\omega)
\int^{\varepsilon_0}_0\, d\varepsilon\,
|G_\omega^z(a,a)|^2
\left(|m_{\omega,+}^z|^2+|m_{\omega,-}^z|^2+2\right)
\;
\frac{\varepsilon}{(C_{9})^2}
\;,
$$
where $\chi_{\Omega_N(\alpha)^c}$ is the indicator function onto the complement of $\Omega_N(\alpha)$. Now one can use the deterministic bound \eqref{eq-bounddeterministic} to conclude that
$$
\langle a|M_q(T)|a\rangle
\; \geq \;
\frac{2\,C_2}{T}
\;\int_0^N dx
\;x^q
\;\int^{\varepsilon_0}_0\, d\varepsilon\;
\frac{\varepsilon}{(C_{9})^2}
\;(1-\PM(\Omega_N(\alpha)))
\;.
$$
Thus there are constants $C_{10}$ and $C_{11}$ such that
$$
\langle a|M_q(T)|a\rangle
\; \geq \;
C_{10}\,\frac{\varepsilon_0^2\,N^{q+1}}{T}
\;=\;
C_{11}\,
T^{-1-\frac{4+8\alpha}{3+2\alpha}+(q+1)\frac{2}{3+\alpha}}
\;.
$$
If $\alpha$ is redefined, this proves Theorem~\ref{theo-transport}.

\end{document}